\documentclass[english,twocolumn,10pt]{IEEEtran}
\usepackage[T1]{fontenc}
\usepackage[latin9]{inputenc}
\usepackage{amsthm}
\usepackage{amsmath}
\usepackage{amssymb}
\usepackage{graphicx}
\usepackage{balance}
\usepackage{epstopdf}
\usepackage{units}
\usepackage[nospace]{cite}
\usepackage{mathrsfs}
\usepackage{algorithm}
\usepackage{relsize}
\usepackage[scr=zapfc,scrscaled=1.2]{mathalfa}
\usepackage[capitalise,nameinlink,noabbrev]{cleveref}

\crefname{equation}{}{}
\crefrangelabelformat{equation}{(#3#1#4)-(#5#2#6)}
\crefname{assumption}{}{} 

\theoremstyle{Lemma}

\DeclareMathAlphabet\mathbfcal{OMS}{cmsy}{b}{n}

\def\T{\mathsf{T}}
\newcommand*{\E}[1]{\mathsf{E} \left\{ #1 \right\}}
\newcommand*{\Ve}[1]{\mathsf{vec} \left\{ #1 \right\}}
\newcommand*{\diag}[1]{\mathsf{diag}\left\{#1\right\}}
\newcommand*{\col}[1]{\mathsf{col}\left\{#1\right\}}
\newcommand*{\p}[1]{\mathsf{p} \left\{ #1 \right\}}
\def \0{\mathbf{0}}

\theoremstyle{Theorem}
\newtheorem{theorem}{Theorem}

\makeatletter

\theoremstyle{plain}
\newtheorem{thm}{\protect\theoremname}
\theoremstyle{remark}
\newtheorem{rem}[thm]{\protect\remarkname}
\newtheorem{assumption}{Assumption}

\makeatother

\providecommand{\remarkname}{Remark}
\providecommand{\theoremname}{Theorem}

\begin{document}

\title{On Stability and Convergence of Distributed Filters}
\author{Sayed Pouria Talebi, Stefan Werner, Vijay Gupta, and Yih-Fang Huang
\thanks{Sayed Pouria Talebi and Stefan Werner are with the Department of Electronic Systems, Faculty of Information Technology and Electrical Engineering, Norwegian University of Science and Technology, Trondheim NO-7491 Norway, E-mail: \{pouria,stefan.werner\}@ntnu.no. 
\\
Vijay Gupta and Yih-Fang Huang are with the Department of Electrical Engineering, University of Notre Dame, Notre Dame, IN 46556 USA, E-mails:\{vgupta2,huang\}@nd.edu.
\\
The work of Sayed Pouria Talebi and Stefan Werner was supported in part by Norwegian Research Council.}} 

\maketitle
\begin{abstract}
Recent years have bore witness to the proliferation of distributed filtering techniques, where a collection of agents communicating over an ad-hoc network aim to collaboratively estimate and track the state of a system. These techniques form the enabling technology of modern multi-agent systems and have gained great importance in the engineering community. Although most distributed filtering techniques come with a set of stability and convergence criteria, the conditions imposed are found to be unnecessarily restrictive. The paradigm of stability and convergence in distributed filtering is revised in this manuscript. Accordingly, a general distributed filter is constructed and its estimation error dynamics is formulated.  The conducted analysis demonstrates that conditions for achieving stable filtering operations are the same as those required in the centralized filtering setting. Finally, the concepts are demonstrated in a Kalman filtering framework and validated using simulation examples. 
\end{abstract}


\section{Introduction}

In recent years, large-scale multi-agent systems that can interact over a network, such as sensor networks, have started to emerge in a wide range of engineering applications including, power system monitoring\cite{PowerSystem,PowerMonitor}, multi-vehicle coordination~\cite{flocking,MultiVehicle}, as well as collaborative information processing and decision making~\cite{CooperativeControl, CentControlTrans,DisOb}. The common thread among most of these applications is the requirement for a robust underlying information processing and filtering framework. Although optimal filtering solutions are attainable through the use of a centralized coordinator, these so-called centralized approaches are vulnerable to failure of the centralized coordinator and require complex communication protocols~\cite{FirstLR,RezaSN,DisFilterControl}. Therefore, one of the most fundamental problems in this area has become that of designing distributed filtering techniques. Distributed filtering techniques aim to enable agents of the network to estimate and track the state of a system or learn a set of parameters in a fashion that limits communication to the neighbourhood of each agent and would not require a centralized coordinator. 

Seminal works in distributed filtering include~\cite{RezaEmbbededConsensus,SayedDiffRLS,RezaSN,SayedDiffLMS,SayedDiffKalman,QFC}, which have since created a vibrant research area (see, e.g.,~\cite{DisControlNet,RobustEstimator,RezaOptimalStable,LinOb,Particle,SayedBook,CooperativeControl,Stefan08,Stefan09,Rev3} and references therein). In general, these distributed filtering techniques are accompanied with stability analysis. However, the assumptions imposed for guaranteeing stable behaviour are, in the authors opinion, restrictive.  For instant, distributed Kalman filtering techniques require the state vector process to be fully observable to each agent via its observation information only or in conjunction with the observation information shared from its neighbours~\cite{SayedDiffKalman,SayedBook,PowerSystem}.  More recent contributions in this field achieve less strict observability criteria, which nonetheless result in the introduction of other requirements that mainly manifest themself as conditions on how information flows throughout the network~\cite{RecentKalman1,RecentKalman2,LinOb}. Despite exhibiting less complex behaviour, the made statements also extend to distributed filters based on gradient descent, e.g., the least mean square (LMS) algorithm~\cite{SayedDiffLMS,DiffLMSPerformance}.

In what follows, a generalized distributed filtering formulation is constructed and its behaviour is analysed. The obtained results  establish that the stability criteria of the considered distributed filter are, in essence, the same as that of its centralized dual. More importantly, due to the generalized setting of the constructed framework, the concept can be extended to a wide range of filtering applications. Finally, the introduced concepts are verified in a Kalman filter setting using simulation examples, showing that in addition to modern techniques~\cite{DisFilterControl,QFC}, traditional distributed Kalman filters~\cite{SayedDiffKalman,RezaSN,SayedBook}  also converge under more relaxed conditions than initially reported.   

\noindent\textit{\textbf{Mathematical Notations}}: Scalars, column vectors, and matrices are denoted by lowercase, bold lowercase, and bold uppercase letters. Inequality symbols refer to generalized matrix inequalities. Remainder of the nomenclature is organized~as~follows:
\begin{IEEEdescription}
\item[$\mathbf{1}$] \hspace{0.2cm}column vector of appropriate size with unit entries
\item[$\mathbf{I}$] \hspace{0.2cm}identity matrix of appropriate size
\item[$(\cdot)^{\T}$] \hspace{0.2cm}transpose operator
\item[$|\cdot|$]\hspace{0.2cm}cardinality operator
\item[$\E{\cdot}$] \hspace{0.2cm}statistical expectation operator
\item[$\p{\cdot}$] \hspace{0.2cm}spectral radius operator
\item[$\Ve{\cdot}$] \hspace{0.2cm}vectorization operator
\item[$\diag{\cdot}$]\hspace{0.2cm}constructs a block-diagonal matrix from its entries  
\item[$\col{\cdot}$]\hspace{0.2cm}constructs a block-column matrix from its entries  
\item[$\otimes$] \hspace{0.2cm}Kronecker product
\item[$\delta(\cdot)$]\hspace{0.2cm}Kronecker delta function
\end{IEEEdescription}

\section{Preliminaries \& Background}

\subsection{Network Model}

Following the conventional setting~\cite{SayedDiffKalman,RezaEmbbededConsensus}, the networked agents are modelled as the connected graph $\mathcal{G}=\{\mathcal{N},\mathcal{E}\}$ with the node set $\mathcal{N}$ representing the agents and their bidirectional communication links represented by the edge set $\mathcal{E}$. The neighbourhood of node $l$ is the set of agents that agent $l$ receives information from, which includes agent $l$ itself, and is denoted by $\mathcal{N}_{l}$. Finally, the cardinality of a node set is the number of nodes that it contains. For example, $|\mathcal{N}_{l}|$ denotes the number of nodes in the neighbourhood of node $l$.

\subsection{Distributed Filtering Paradigm}

Consider the state vector  sequence $\{\mathbf{x}_{n},n=1,2,\ldots\}$ described through the linear dynamics
\begin{equation}
\mathbf{x}_{n+1}=\mathbf{A}\mathbf{x}_{n}+\mathbf{v}_{n}
\label{eq:StateVector}
\end{equation}
with $\mathbf{A}$ denoting the state evolution matrix and $\mathbf{v}_{n}$ representing the state evolution noise at time $n$. The state vector is observed at node $l$, so that we have
\begin{equation}
\mathbf{y}_{l,n}=\mathbf{H}_{l,n}\mathbf{x}_{n}+\mathbf{w}_{l,n}
\label{eq:Observation}
\end{equation}
where at time instant $n$ and node $l$, $\mathbf{y}_{l,n}$ is the observation and $\mathbf{H}_{l,n}$ is the observation  matrix, while $\mathbf{w}_{l,n}$ denotes the observation noise.

\begin{assumption}
In keeping with the conventional distributed filtering framework~\cite{SayedDiffKalman,RezaEmbbededConsensus,RezaSN,DisFilterControl}, all noise sequences are assumed to be stationary zero-mean Gaussian processes, with covariance matrix 
\[
\E{\begin{bmatrix}\mathbf{v}_{n} \\ \mathbf{w}_{l,n} \end{bmatrix}\begin{bmatrix}\mathbf{v}^{\T}_{m} & \mathbf{w}^{\T}_{i,m}
\end{bmatrix}} = \begin{bmatrix} \boldsymbol{\Sigma}_{\mathbf{v}} & \mathbf{0} \\ \mathbf{0} &
\boldsymbol{\Sigma}_{\mathbf{w}_{l}}\delta(l-i)\end{bmatrix}\delta(n-m).
\]
\label{Assumption:Noise} 
\end{assumption}

A centralized filtering operation can be achieved through the following operations~\cite{DisFilterControl,RezaEmbbededConsensus}
\begin{equation}
\hat{\mathbf{x}}_{n+1}=\left(\mathbf{I}-\mathbf{G}\mathbf{H}_{\mathrm{col}}\right)\hat{\mathbf{x}}_{n}+\mathbf{G}\mathbf{y}_{\mathrm{col},n}
\label{eq:CentralizedFilter}
\end{equation}
where $\mathbf{G}$ is an adaptation gain, whereas 
\[
\mathbf{H}_{\mathrm{col}}=\col{\mathbf{H}_{l}:\forall l\in\mathcal{N}}\hspace{0.2cm}\text{and}\hspace{0.2cm}\mathbf{y}_{\mathrm{col},n}=\col{\mathbf{y}_{l,n}:\forall l\in\mathcal{N}}.
\]
Importantly, if $\{\mathbf{A},\mathbf{H}_{\mathrm{col}}\}$ is detectable and $\{\mathbf{A},\boldsymbol{\Sigma}^{\frac{1}{2}}_{\mathbf{v}}\}$ is stabilizable; then,  the algebraic Riccati recursion 
\begin{equation}
\mathbf{M}^{{}^{-1}}_{n+1}=\left(\mathbf{A}\mathbf{M}_{n}\mathbf{A}^{\T}+\boldsymbol{\Sigma}_{\mathbf{v}}\right)^{{}^{-1}}+\mathbf{H}^{\T}_{\mathrm{col}}\mathbf{R}^{-1}\mathbf{H}_{\mathrm{col}}
\label{eq:CentralRiccati}
\end{equation}
where $\mathbf{R}$ is a block-diagonal weighting matrix given as
\[
\mathbf{R}=\diag{\mathbf{R}_{l}:l=1,\ldots,|\mathbf{N}|}\hspace{0.12cm}\text{so~that} \hspace{0.12cm}\forall l:\mathbf{R}_{l}\geq 0
\]
converges to a unique stabilizing solution $\mathbf{M}$, so that  
\[
\forall \mathbf{M}_{1}>0:\hspace{0.12cm}\lim_{n\rightarrow\infty}\mathbf{M}_{n}=\mathbf{M}
\] 
for which the  adaptation gain $\mathbf{G}=\mathbf{M}\mathbf{H}^{\T}_{\mathrm{col}}\mathbf{R}^{-1}$ makes the estimation error of the filtering operations in \eqref{eq:CentralizedFilter} into a globally stable linear system~\cite{LinEst,AdaptiveControl}.

The aim in distributed filtering is to enable each node in the network to track the state vector sequence in \eqref{eq:StateVector} via observations available to the agent as described in \eqref{eq:Observation} and cooperation achievable through communication only with its neighbouring nodes. In the sequel, a general framework for achieving this goal is introduced. Then, its behaviour is rigorously analysed, establishing its stable operating criteria. Finally, the results are shown to be applicable to a number of seminal distributed filtering approaches.

\section{Distributed Filtering Framework}

\subsection{Distributed Filtering Operations}

To achieve a distributed filtering operation, and without loss of generality,  consider the following operations at each node 
\begin{equation}
\forall l\in\mathcal{N}:\left\{\begin{aligned}\boldsymbol{\phi}_{l,n+1}=&\left(\mathbf{I}-\mathbf{G}_{l}\mathbf{H}_{l}\right)\mathbf{A}\hat{\mathbf{x}}_{l,n}+\mathbf{G}_{l}\mathbf{y}_{l,n}
\\
\hat{\mathbf{x}}_{l,n+1}=&\sum_{\forall i\in\mathcal{N}_{l}}c_{l,i}\boldsymbol{\phi}_{i,n+1}
\end{aligned}
\right.
\label{eq:Filter}
\end{equation}
where node $l$ implements a local filtering operation using adaptation gain $\mathbf{G}_{l}$; then, combines its intermediate estimate, $\boldsymbol{\phi}_{l,n}$, with those of its neighbours using the combination weights $\{c_{l,i}:\forall l,i\in\mathcal{N}\}$ to arrive at the final estimate $\hat{\mathbf{x}}_{l,n}$. 

\begin{rem}
The filtering operation in \eqref{eq:Filter} is designed to encompass the seminal works in distributed Kalman filtering~\cite{SayedDiffKalman,RezaOptimalStable,DisFilterControl}. In addition, it is straightforward to simplify the operation in \eqref{eq:Filter} to describe distributed filters based on gradient descent~\cite{SayedBook,SayedDiffLMS,DiffLMSPerformance}.
\end{rem}

\begin{assumption}
In keeping with the diffusion and consensus information fusion setting~\cite{AverageConsensusJournal,AverageConsensus,RezaConsensus,SayedBook}, the combination weights in \eqref{eq:Filter} are chosen so that
\[
\forall l\in\mathcal{N}:\sum_{\forall i\in\mathcal{N}_{l}}c_{l,i}=1.
\]
Thus, matrix $\mathbf{C}$ with $l^{\text{th}}$ row and $i^{\text{th}}$ column element 
\[
\mathbf{C}^{\{l,i\}}=\left\{\begin{matrix}c_{l,i}&{}&\text{if}\hspace{0.12cm}i\in\mathcal{N}_{l}\\0&{}&\text{if}\hspace{0.12cm}i\notin\mathcal{N}_{l}\end{matrix}\right.
\]
would be right-stochastic and primitive~\cite{SayedBook}.  
\label{Assumption:Affine}
\end{assumption}

\subsection{Stability and Convergence Analysis}

The filtering operations in \eqref{eq:Filter} are combined to give 
\begin{equation}
\hat{\mathbf{x}}_{l,n+1}=\sum_{\forall i\in\mathcal{N}_{l}}c_{l,i}\Big(\left(\mathbf{I}-\mathbf{G}_{i}\mathbf{H}_{i}\right)\mathbf{A}\hat{\mathbf{x}}_{i,n}+\mathbf{G}_{i}\mathbf{y}_{i,n}\Big).
\label{eq:FilterCombined}
\end{equation}
Now, subtracting $\mathbf{x}_{l,n+1}$ from both sides of \eqref{eq:FilterCombined} yields
\begin{align}
\boldsymbol{\epsilon}_{l,n+1}=&\mathbf{x}_{n+1}-\hat{\mathbf{x}}_{l,n+1}\label{eq:FirstError}
\\
=&\mathbf{x}_{n+1}-\sum_{\forall i\in\mathcal{N}_{l}}c_{l,i}\Big(\left(\mathbf{I}-\mathbf{G}_{i}\mathbf{H}_{i}\right)\mathbf{A}\hat{\mathbf{x}}_{i,n}+\mathbf{G}_{i}\mathbf{y}_{i,n}\Big).\nonumber
\end{align}
Substituting $\mathbf{x}_{n+1}$ from \eqref{eq:StateVector} and $\mathbf{y}_{l,n}$ from \eqref{eq:Observation} into \eqref{eq:FirstError} gives the regressive error expression
\begin{align}
\boldsymbol{\epsilon}_{l,n+1}=&\sum_{\forall i\in\mathcal{N}_{l}}c_{l,i}\left(\mathbf{I}-\mathbf{G}_{i}\mathbf{H}_{i}\right)\mathbf{A}\boldsymbol{\epsilon}_{i,n}-\sum_{\forall i\in\mathcal{N}_{l}}c_{l,i}\mathbf{G}_{i}\mathbf{w}_{i,n}\nonumber
\\
&+\sum_{\forall i\in\mathcal{N}_{l}}c_{l,i}\left(\mathbf{I}-\mathbf{G}_{i}\mathbf{H}_{i}\right)\mathbf{v}_{n}.\label{eq:ErrorReg}
\end{align}

The expression in \eqref{eq:ErrorReg} shows how the performance of node $l$ is linked to that of its neighbours, which are in turn, linked to their respective neighbours. Therefore, an over all analysis is only possible if the entire network is considered as unified filter. To this end, the network-wide error vector is defined as
\begin{equation}
\mathbfcal{E}_{n}=\col{\boldsymbol{\epsilon}_{l,n}:\forall l\in\mathcal{N}}.
\label{eq:NetErrorDef}
\end{equation}
After some mathematical manipulation, replacing the rows of $\mathbfcal{E}_{n}$ in \eqref{eq:NetErrorDef} with expressions in \eqref{eq:ErrorReg} gives 
\begin{equation}
\mathbfcal{E}_{n+1}=\mathbfcal{C}\Big(\mathbfcal{F}\mathbfcal{E}_{n}+\mathbfcal{P}\left(\mathbf{1}\otimes\mathbf{v}_{n}\right)-\mathbfcal{G}\mathbfcal{W}_{n}\Big)
\label{eq:FianlErrorRecurtion}
\end{equation}
where $\mathbfcal{C}=\mathbf{C}\otimes\mathbf{I}$, $\mathbfcal{G}=\text{diag}\{\mathbf{G}_{l}:\forall l\in\mathcal{N}\}$, and
\begin{align}
\mathbfcal{F}=&\diag{\left(\mathbf{I}-\mathbf{G}_{l}\mathbf{H}_{l}\right)\mathbf{A}:\forall l\in\mathcal{N}}\label{eq:DisErrorGain}
\\
\mathbfcal{P}=&\diag{\left(\mathbf{I}-\mathbf{G}_{l}\mathbf{H}_{l}\right):\forall l\in\mathcal{N}}
\\
\mathbfcal{W}_{n}=&\col{\mathbf{w}_{l,n}:\forall l \in\mathcal{N}}.
\end{align}

\begin{theorem}
If $\{\mathbf{A},\mathbf{H}_{\mathrm{col}}\}$ is detectable and $\{\mathbf{A},\boldsymbol{\Sigma}^{\frac{1}{2}}_{\mathbf{v}}\}$ is stabilizable; then, there exists a matrix set $\{\mathbf{M}_{l}:\forall l\in\mathcal{N}\}$ so that
\begin{equation}
\forall l\in\mathcal{N}:\mathbf{G}_{l}=\mathbf{M}_{l}\mathbf{H}^{\T}_{l}\mathbf{R}^{{}^{-1}}_{l}
\label{eq:DisGain}
\end{equation}
makes the estimation error of the distributed filtering operations in \eqref{eq:Filter} globally stable.
\label{Theorem:Stability}
\end{theorem}

\begin{IEEEproof}[Proof of Theorem~\ref{Theorem:Stability}]
Take the matrix recursions
\begin{subequations}
\begin{align}
\mathbf{S}_{l,n+1} =&\left(\mathbf{A}\mathbf{M}_{l,n}\mathbf{A}^{\T}+\boldsymbol{\Sigma}_{\mathbf{v}}\right)^{{}^{-1}}+\mathbf{H}^{\T}_{l}\mathbf{R}^{{}^{-1}}_{l}\mathbf{H}_{l}
\\
\mathbf{M}^{{}^{-1}}_{l,n+1}=&\sum_{\forall l\in\mathcal{N}_{l}}c_{l,i}\mathbf{S}_{i,n+1}
\end{align}
\label{eq:DisRiccati}
\end{subequations}
which can be carried out by the agents in a distributed fashion. Comparing \eqref{eq:DisRiccati}  to \eqref{eq:CentralRiccati}, and assuming, without loss of generality, that $\forall l\in\mathcal{N}:0<\mathbf{M}_{1}\leq\mathbf{M}_{l,1}$, it can be concluded that $\forall l\in\mathcal{N}:0<\mathbf{M}_{n}\leq\mathbf{M}_{l,n}$. Proceeding on this basis, it follows algebraically that for $\forall l\in\mathcal{N}$:
\begin{equation}
\left(\mathbf{I}-\mathbf{M}_{l,n}\mathbf{H}^{\T}_{l}\mathbf{R}^{{}^{-1}}_{l}\mathbf{H}_{l}\right)\leq\left(\mathbf{I}-\mathbf{M}_{n}\mathbf{H}^{\T}_{l}\mathbf{R}^{{}^{-1}}_{l}\mathbf{H}_{l}\right).
\label{eq:StabilityRiccatiFirst}
\end{equation}
Moreover, the assumption that $\{\mathbf{A},\mathbf{H}_{\mathrm{col}}\}$ is detectable and $\{\mathbf{A},\boldsymbol{\Sigma}^{\frac{1}{2}}_{\mathbf{v}}\}$ is stabilizable, from the Riccati recursions in \eqref{eq:CentralRiccati}, we have
\begin{equation}
\p{\left(\mathbf{I}-\mathbf{M}_{n}\sum_{\forall l\in\mathcal{N}}\mathbf{H}^{\T}_{l}\mathbf{R}^{{}^{-1}}_{l}\mathbf{H}_{l}\right)\mathbf{A}}<1.
\label{eq:CentStable}
\end{equation}
Thus, from  \eqref{eq:CentStable} and \eqref{eq:StabilityRiccatiFirst} it follows that, $(\mathbf{I}-\mathbf{M}_{l,n}\mathbf{H}^{\T}_{l}\mathbf{R}^{{}^{-1}}_{l}\mathbf{H}_{l})\mathbf{A}$, would be stable on modes detectable at node $l$. Subsequently, $\sum_{\forall l\in\mathcal{N}_{l}}c_{l,i}\left(\mathbf{I}-\mathbf{M}_{i,n}\mathbf{H}^{\T}_{i}\mathbf{R}^{{}^{-1}}_{i}\mathbf{H}_{i}\right)\mathbf{A}$, would be stable on modes detectable to nodes in $\mathcal{N}_{l}$. 

From Assumption~\ref{Assumption:Affine}, recall that $\mathbf{C}$ is primitive, and hence $\mathbfcal{C}$ is block primitive, that is, there exits $m$ so that $\mathbfcal{C}^{m}=\mathbf{C}^{m}\otimes\mathbf{I}$ consists of identity matrices scaled by positive non-zero real-valued numbers.  Given the block diagonal structure of $\mathbfcal{F}$, extending the statements in the previous paragraph, there exists $k$ for which $\left(\mathbfcal{C}\mathbfcal{F}\right)^{k}$ is a block matrix, where each block consists of an appropriate combination of the matrix set
\[
\left\{\left(\left(\mathbf{I}-\mathbf{M}_{l,n}\mathbf{H}^{\T}_{l}\mathbf{R}^{{}^{-1}}_{l}\mathbf{H}_{l}\right)\mathbf{A}\right)^{i}:\forall l\in\mathcal{N},i=1,\ldots,k\right\}
\] 
so that $\p{(\mathbfcal{C}\mathbfcal{F})^{k}}<1$. Therefore, the statistical expectation of any norm, including the second-order norm, of the estimation error sequence 
$\{\mathbfcal{E}_{n+k}:n=1,2,\dots\}$ would become convergent, which concludes the proof. 
\end{IEEEproof}

\begin{rem}
Note that Theorem~\ref{Theorem:Stability} broadens the scope of our previous results in~\cite{DisFilterControl} and presents the least restrictive convergence criteria for an all-inclusive class of distributed filters, essentially equating convergence criteria of centralized and distributed filtering techniques.
\end{rem}

\begin{theorem}
If $\{\mathbf{A},\mathbf{H}_{\mathrm{col}}\}$ is detectable and $\{\mathbf{A},\boldsymbol{\Sigma}^{\frac{1}{2}}_{\mathbf{v}}\}$ is stabilizable; then, the gain matrices resulting from \eqref{eq:DisGain} and \eqref{eq:DisRiccati} make the estimates of the introduced distributed filtering operations in \eqref{eq:Filter} globally asymptotically unbiased.
\label{Theorem:Bias}
\end{theorem}

\begin{IEEEproof}[Proof of Theorem~\ref{Theorem:Bias}]
From the recursive expression of the estimation error in \eqref{eq:FianlErrorRecurtion} and conditions set on state evolution and observation noise in Assumption~\ref{Assumption:Noise}, we have
\begin{equation}
\E{\mathbfcal{E}_{n}}=\left(\mathbfcal{C}\mathbfcal{F}\right)^{n}\E{\mathbfcal{E}_{1}}.
\label{eq:ErrorBias}
\end{equation}
Furthermore, in the proof of Theorem~\ref{Theorem:Stability}, it was demonstrated that under the made assumptions, there exits a $k$ for which $\left(\mathbfcal{C}\mathbfcal{F}\right)^{k}$ is a contracting operator, and therefore,
\begin{equation}
\text{as}\hspace{0.12cm}n\rightarrow\infty\hspace{0.12cm}\text{then}\hspace{0.12cm}\left(\mathbfcal{C}\mathbfcal{F}\right)^{n}\rightarrow\mathbf{0}.
\label{eq:CFinf}
\end{equation}
As a direct result of \eqref{eq:ErrorBias} and \eqref{eq:CFinf}, we have
\[
\lim_{n\rightarrow\infty}\E{\mathbfcal{E}_{n}}=\mathbf{0}
\]
which concludes the proof.
\end{IEEEproof}

To provide a more practical perspective, a reformulation of \eqref{eq:DisRiccati} is considered, so that
\begin{equation}
\mathbf{M}^{{}^{-1}}_{l,n}=\sum_{\forall i\in\mathcal{N}_{l}}c_{l,i}\mathbf{S}_{i,n}=\sum_{\forall i\in\mathcal{N}_{l}}c_{l,i}\mathbf{P}^{{}^{-1}}_{i,n}+\bar{\mathbf{H}}^{\T}_{l}\mathbf{R}^{{}^{-1}}\bar{\mathbf{H}}_{l}
\label{eq:DisRiccatiExpantion}
\end{equation}
where $\forall l\in\mathcal{N}:\mathbf{P}_{l,n} =\mathbf{A}\mathbf{M}_{l,n-1}\mathbf{A}^{\T}+\boldsymbol{\Sigma}_{\mathbf{v}}$, while
\[
\bar{\mathbf{H}}_{l}=\left[\sqrt{\mathbf{C}^{\{l,1\}}}\mathbf{H}^{\T}_{1},\ldots,\sqrt{\mathbf{C}^{\{l,|\mathcal{N}|\}}}\mathbf{H}^{\T}_{|\mathcal{N}|}\right]^{\T}.
\]
Now, using \eqref{eq:DisRiccatiExpantion} and some mathematical manipulation, the operations in \eqref{eq:DisRiccati} are rearranged as
\begin{subequations}
\begin{align}
\mathbf{P}_{l,n}=&\mathbf{A}\left(\mathbf{Z}^{{}^{-1}}_{l,n-1}+\bar{\mathbf{H}}^{\T}_{l}\mathbf{R}^{{}^{-1}}\bar{\mathbf{H}}_{l}\right)^{{}^{-1}}\mathbf{A}^{\T}+\boldsymbol{\Sigma}_{\mathbf{v}}\label{eq:DistRiccatiState}
\\
\mathbf{Z}^{{}^{-1}}_{l,n}=&\sum_{\forall i\in\mathcal{N}_{l}}c_{l,i}\mathbf{P}^{{}^{-1}}_{i,n}.\label{eq:DistRiccatiCombine}
\end{align}
\label{eq:DistRiccati}
\end{subequations} 

From \eqref{eq:DistRiccatiState}, it is clear that $\mathbf{P}_{l,n}$ is dual of the \textit{a~posteriori} estimate error covariance matrix of a Kalman filter with \textit{a~priori} estimate error covariance of $\mathbf{Z}_{l,n-1}$ using observations $\{\mathbf{y}_{i,n}:\forall i\in\mathcal{N}_{l}\}$. Following on the same line, from \eqref{eq:DistRiccatiCombine}, $\mathbf{Z}_{l,n}$ becomes the dual of the post estimate error covariance of a Kalman filter using observations $\{\mathbf{y}_{i,n}:\forall i\in\mathcal{N}_{l_{2}}\}$, where $\mathcal{N}_{l_{k}}$ is  the $k$-hop neighbourhood of node $l$.  Thus, through iteration, $\mathbf{Z}_{l,n+m}$ becomes the dual of the \textit{a~posteriori} estimate error covariance of a Kalman filter using observations
\begin{equation}
\bigcup\limits^{m}_{t=0}\{\mathbf{y}_{i,n+k}:\forall i\in\mathcal{N}_{l_{t+1}},k=0,\ldots,m-t\}.
\label{eq:DistObs}
\end{equation}
If $\{\mathbf{A},\mathbf{H}_{\mathrm{col}}\}$ is detectable and $\{\mathbf{A},\boldsymbol{\Sigma}^{\frac{1}{2}}_{\mathbf{v}}\}$ is stabilizable; then, there exists $m$ for which observation set in \eqref{eq:DistObs} is sufficient for tracking the state vector, and thus, $\{\mathbf{M}_{l,n}:\forall l\in\mathcal{N}\}$ converges to stabilizing solutions.

It is prudent to note that the Riccati style recursions in \eqref{eq:DisRiccati} have their roots in the decentralized controllers in~\cite{DisFilterControl,QFC}. The structure in~\eqref{eq:DisRiccati}  is specifically designed to stabilize controllers with the dynamics in~\eqref{eq:FianlErrorRecurtion}, which results in its versatile application in this context. However, this manuscript, further eases convergence criteria and removes the restriction for using the consensus framework for information fusion, accommodating for both diffusion and consensus. Traditional distributed Kalman filters inspired from~\cite{SayedDiffKalman,RezaSN}  that do not use~\eqref{eq:DisRiccati} and are reliant on local observability conditions, will experience unstable modes in their corresponding $\mathbf{M}_{l}$ matrices. However, even in these cases, if the error dynamic follows \eqref{eq:FianlErrorRecurtion} and the gain follows \eqref{eq:DisGain}, the overall filtering operation will converge. Note that this might take some skilful coding legerdemain to ensure unstable modes do not result in computation errors, which fall beyond the scope of this manuscript. On a final note, the distributed filter in \eqref{eq:Filter} can be simplified to the distributed filters based on the  LMS~\cite{SayedDiffLMS,DiffLMSPerformance,SayedBook} and even a wide range of gradient-based approaches for nonlinear filters, in which case, the concepts introduced in this manuscript would hold true. 

\section{Simulation Examples}

\begin{figure}[H]
\centering
\includegraphics[width=0.48\linewidth]{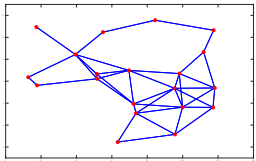}
\caption{Network of $20$ nodes and $40$ links used in simulations.}
\label{fig:network}
\end{figure}

The problem of tracking a target in two dimensions is considered~\cite{SayedDiffKalman,DisFilterControl}. In this case, the state evolution function is
\[
\mathbf{x}_{n} = \begin{bmatrix}1&0&\Delta T&0\\0&1&0&\Delta T\\0&0&1&0\\0&0&0&1\end{bmatrix}\mathbf{x}_{n-1} + \begin{bmatrix}\frac{1}{2}(\Delta T)^{2}&0\\0&\frac{1}{2}(\Delta T)^{2}\\\Delta T&0\\0&\Delta T\end{bmatrix}\mathbf{v}_{n} \\
\]
where $\E{\mathbf{v}_{n}\mathbf{v}^{\T}_{n}}=0.01\times\mathbf{I}$ and $\Delta T=$\unit[$0.04$]{s} representing the sampling interval. The network shown in Fig.~\ref{fig:network} was used for simulations. Excluding the node with one link sitting on the top left, all nodes observed movements of the target on the horizontal axis solely. The mentioned node, on the other hand, could only observe moments on the vertical axis, presenting a worst-case scenario for distributed filtering. The observation noise covariance for all nodes was $0.16$ which was used as the weight factors $\mathbf{R}_{l}$.

\begin{figure}[H]
\centering
\includegraphics[width=0.92\linewidth, trim = 2cm 2cm 2cm 2.4cm]{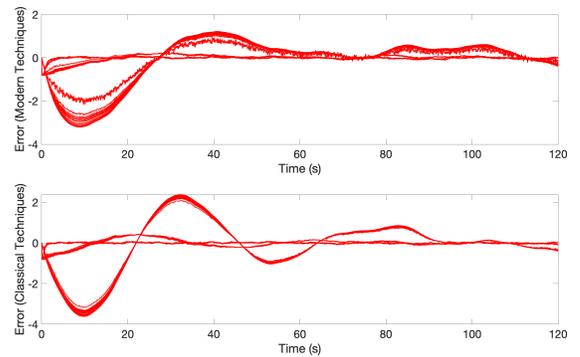}
\caption{Error dynamics for both modern and classical distributed Kalman filtering techniques.}
\label{Fig:Error}
\end{figure}

The distributed Kalman filter in~\cite{SayedDiffKalman}, as a representative of classical techniques, and the distributed Kalman filter in~\cite{DisFilterControl}, as a representative of modern techniques, were used to track the state vector. The estimation error dynamics for both distributed filters is presented in Fig.~\ref{Fig:Error}. Note that, as suggested in Theorem~\ref{Theorem:Stability}~and~\ref{Theorem:Bias}, both distributed Kalman filtering approaches exhibited error dynamics that converge to zero. This result is specially intriguing in the case of traditional distributed Kalman filtering techniques as the scenario considered here does not meet their original convergence criteria (see~\cite{SayedDiffKalman,SayedBook}).    

\begin{figure}[H]
\centering
\includegraphics[width=0.92\linewidth, trim = 2cm 2cm 2cm 3.4cm]{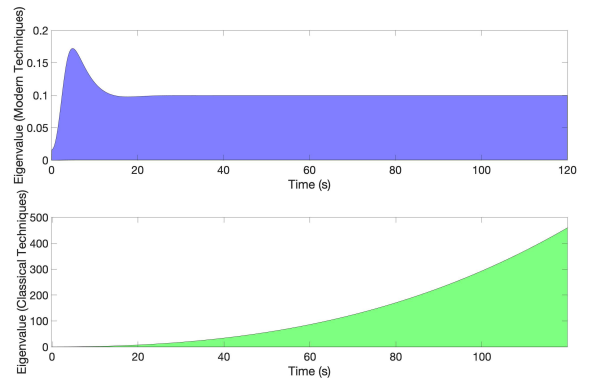}
\caption{Range of eigenvalues for the matrix set $\{\mathbf{M}_{l,n}:\forall l\in\mathcal{N}\}$ for both modern and classical distributed Kalman filtering techniques.}
\label{Fig:EigVal}
\end{figure}

The range for eigenvalues of matrix set $\{\mathbf{M}_{l,n}:\forall l\in\mathcal{N}\}$ obtained through the recursions in \eqref{eq:DisRiccati} are presented in Fig.~\ref{Fig:EigVal}. In addition, Fig.~\ref{Fig:EigVal} includes the range of eigenvalues of comparable set of matrices obtained through the framework of~\cite{SayedDiffKalman,RezaSN,SayedBook}. Note that as proven in Theorem~\ref{Theorem:Stability}, the eigenvalues of matrices in \eqref{eq:DisRiccati} converge and stabilize to a final value, whereas their counterparts from classical filtering frameworks diverge. It can be shown through straightforward algebraic manipulations that the divergent modes at each node correspond to modes unobservable at that node, which in principle, results in these modes only being updated through the state evolution model. This will not cause a divergent behaviour, however, as information regarding modes unobservable to one node will be made available through the diffusion/consensus structure as long as there exists another node in the network to whom these modes are observable, which encapsulates the spirit of this manuscript in a more intuitive manner. It should be noted that when using classical approaches, the presence of the unobservable modes for each node should be considered in the implementation phase, as the infinitely growing eigenvalues can cause computational problems. However, modern approaches, such as~\cite{DisFilterControl,QFC} and that presented in this manuscript, deal with this matter through rigorous modelling and skilful use of mathematics.

\section{Conclusion}

The performance of a general class of distributed filters has been considered. The analysis culminated in two theorems, which indicate that distributed filters and their centralized duals essentially require the same convergence criteria. The distributed filters considered were formulated to make the obtained results applicable in a wide range of distributed filtering frameworks.  The concepts were verified using simulation examples on both modern and classical distributed Kalman filters, showing convergence is possible under much looser conditions than previously thought.

\balance
 
\bibliographystyle{IEEEtran}

\bibliography{ref}

\end{document}